\documentstyle[preprint,prb,aps]{revtex}
\title{The Abrikosov Flux Lattice in Planar Crystals of
YBa$_{2}$Cu$_{3}$O$_{7-\delta}$}
\author{M. Harrison and W. Barford$^*$}
\address{Department of Physics,
The University of Sheffield,\\ Sheffield, S3 7RH, United Kingdom.}
\date{\today}
\begin{document}
\maketitle
\draft
\begin{abstract}

Anisotropic London theory is used to predict the Abrikosov flux
lattice for arbitrary field orientations in crystals of
YBa$_{2}$Cu$_{3}$O$_{7-\delta}$ by minimising the Gibbs free energy
for samples of a planar geometry. At low fields the ``vortex-chain''
state exists, {\it i.e.} the inter-chain vortex separation scales as
the inverse flux density, $1/B$, whereas the intra-chain vortex
separation is constant.  At higher fields there is a cross-over to the
uniaxially distorted hexagonal lattice where the inter-vortex
separation scales as $1/B^{1/2}$. At low fields the vortex lattice is
inclined towards the $\hat{\bf c}$ axis, rotating towards the applied
field at higher fields. The results of this calculation are in close
agreement to the Bitter pattern experiments of Gammel {\it et
al.}\cite{gammel92} assumming an anisotropy ratio $\gamma$ of 5 and an
in-plane penetration depth $\lambda_a$ of 1 413 \AA, indicating that
the ``vortex-chain'' state has been observed in these experiments.

\end{abstract}

\vspace {1 truein}

* To whom correspondence should be addressed.

\pagebreak

\section{Introduction}
An interesting aspect of high temperature superconductors is the
nature of the mixed state in which magnetic flux lines (or vortices)
penetrate the superconducting sample when the applied field exceeds
H$_{c1}$.  In isotropic materials the vortices are arranged in a
regular hexagonal lattice \cite{ab57}. In the anisotropic high
temperature superconductors, however, the vortex lattice is distorted
for an applied field not parallel to the crystalline axis. The
detailed structure and arrangement of vortices in these materials has
been the subject of much experimental and theoretical attention in the
last few years.

For high magnetic fields the distortion of the equilibrium flux line
lattice depends on the anisotropy ratio and orientation of the lattice
with respect to the crystalline axis\cite{kog88}.  At low magnetic
fields, however, it has been predicted that the flux line lattice
becomes distorted into a series of equally spaced vortex chains,
running parallel to the plane containing the applied field and the
$\hat{\bf c}$ axis: the $\theta$ plane.  This ``vortex-chain''
structure arises from the vortex-vortex interaction becoming
attractive for magnetic fields not parallel to one of the principle
crystalographic axes \cite{grishin,kog90}.  The minima of the
interaction are situated at a distance $x_{min}$ on either side of the
vortex core in the $\theta$ plane. The lattice is no longer expected
to scale uniformly with field; the intra-chain spacing, the distance
between vortices within a chain, is determined by $x_{min}$. For a
fixed magnetic flux density the intra-vortex spacing remains constant,
and hence the inter-vortex spacing, the distance between chains, is
inversely proportional to the flux density.

A number of experiments have investigated the flux lattice directly;
both on the surface of the sample, and within the bulk. Small angle
neutron scattering experiments have been performed to probe the
magnetic field distribution within single crystals at high
fields. Yethriaj {\it et al.} \cite{yethiraj93} observed a distorted
hexagonal lattice in YBa$_{2}$Cu$_{3}$O$_{7-\delta}$, in agreement
with the predictions of London theory.  Keimer {\it et
al}.\cite{keimer}, however, observed vortex-chains for an applied
field of $B=0.5T$ inclined at 80$^{o}$ to the $\hat {\bf c}$ axis.
The presence of vortex-chains at these large fields is attributed to
an exponential softening of the vortex lattice shear modulus, as
predicted by Ivlev and Kopnin\cite{ivlev}.

High-resolution Bitter decoration experiments have also been performed
to directly observe the flux line lattice emerging from the surface of
single, twin free crystals of YBa$_{2}$Cu$_{3}$O$_{7-\delta}$ in low
fields by Gammel and {\it et al.}\cite{gammel92}. Their results show
the variation of the intra- and inter- chain distance as a function of
the normal component of the applied field, $H_{Z}$.  These results
indicate that both the intra- and inter- chain distances scale as
roughly $1/{\sqrt H_Z}$, and hence do not appear to obey the
vortex-chain state predictions.  However, since the experiments are
performed on flat, platelet crystals, demagnetisation effects play an
important r\^ole in determining the density and orientation of the
flux lattice with respect to the crystalline axis. It is the purpose
of this paper to show that once the demagnetisation effects have been
considered the experimental results fit the vortex-chain predictions
very well.  This implies that, although the Bitter decoration
experiments are surface probes, they do provide useful information on
the bulk properties of the lattice. It also demonstrates that three
dimensional anisotropic London theory provides a good description of
the flux lattice in YBa$_{2}$Cu$_{3}$O$_{7-\delta}$.

The plan of this paper is as follows: in \S II we discuss the
thermodynamics of planar superconductors in a magnetic field.  This is
the situation relevant to most of the recent decoration experiments on
high temperature superconductors. \S III discusses the magnetic field
distributions obtained from anisotropic London theory, which enables
the pairwise interaction between vortices to be calculated.  In \S IV
we explain the details of the lattice summation, for both the low and
high field limits, while \S V describes the results and a comparison
with the experimental observations. We summarise in \S VI.

\section{The Thermodynamics of Planar Superconductors}

The high resolution Bitter pattern experiments of Gammel {\it et al.}
\cite{gammel92}were performed on flat platelets approximately 0.5mm x
0.5mm square, and between 5 to 40 $\mu$m in thickness. Thus,
demagnetisation effects must be considered in the determination of the
flux line lattice.  In figure 1 we illustrate the coordinate system
used in this paper.  The ($\hat{\bf X},\hat{\bf Y},\hat{\bf Z}$) axes
define the crystallographic axes with $\hat{\bf Z}$ being parallel to
$\hat{\bf c}$ and ($\hat{\bf X} - \hat{\bf Y}$) lying in the
($\hat{\bf a}-\hat{\bf b}$) plane.  In general an applied field, ${\bf
H}_a$, will be oriented at an angle $\phi$ with respect to the
$\hat{\bf c}$ axis.  Since we are only concerned with uniaxial
anisotropy, the plane enclosed by $\hat{\bf c}$ and ${\bf H}_a$ is a
symmetry plane.  Hence, the magnetic flux density, $\bf B$, defined as
the average flux density per unit cell, also lies in this symmetry
plane, at an angle $\theta$ with respect to the $\hat{\bf c}$ axis. We
therefore define a coordinate axes ($\hat{\bf x},\hat{\bf y},\hat{\bf
z}$) by a rotation of $\theta$ about the $\hat{\bf Y}$ axis, such that
$\hat {\bf z}$ is parallel to $\bf B$.

 To simplify matters we will assume that the platelets are infinite in
the $\hat{\bf a}-
\hat{\bf b}$
($\hat{\bf X}-\hat{\bf Y}$)
plane. The boundary conditions are then that the normal component of
{\bf B} and the parallel component
of {\bf H} to the surfaces are continuous, {\it i.e.},
\begin{equation}
B_{Z}=B_{a} \cos\phi  =B\cos\theta
\end{equation}
and
\begin{equation}
H_{X}=H_{a} \sin\phi,
\end{equation}
where $B_a=H_a$.
Equation (1) is generally satisfied in Bitter pattern experiments on
flat platelets; see for example, figure (1) in reference
[\onlinecite{gammel92}]. The flux density is determined by,
\begin{equation}
B=\frac{B_{a} \cos\phi}{\cos\theta}.
\end{equation}

In the presence of fixed external currents and temperature the
equilibrium flux line lattice is obtained by minimising the Gibbs free
energy,
\begin{equation}
G({\bf H})=U-\frac{{\bf B.H}}{4\pi},
\end{equation}
where $U$ contains the interaction between the flux lines and their
self-energies, and we have defined G in CGS units.

 The Gibbs energy is stationary with respect to the normal component
 of {\bf B}, and hence the correct thermodynamic potential to be
 minimised is,
\begin{equation}
G=U-\frac{B_{X}H_{X}}{4\pi}
\end{equation}
where $H_{X}$ is given by equation (2) and $B_{X}$ by
\begin{equation}
B_{X}=B \sin\theta.
\end{equation}

As shown by Kogan\cite{Vkogan90}, the general expression for the
pairwise interaction per unit length, $U_{12}$, between two parallel
vortices is
\begin{equation}
U_{12}=\frac{\phi_{o}}{4\pi}h_{z}({\bf r}_{12})
\end{equation}
where $h_{z}({\bf r_{12}})$ is the longitudinal component of the
magnetic field due to vortex $2$ at the position of vortex $1$ (and
{\it vice versa}).  In a periodic lattice all points are
equivalent. Hence, the lattice sum is obtained by summing the pairwise
interactions between a vortex at the origin and another at the lattice
position {\bf R}$_{i}$, multiplied by the vortex number density, $n$.
Then the interaction energy per unit volume is,
\begin{equation}
U=\frac{n\phi_{o}}{8\pi}\sum_{{\bf R}_{i}}h_{z}\left({\bf R}_{i}\right).
\end{equation}
This sum includes the self-energy of the vortices, taken as the limit
${\bf R}_{i}\rightarrow0$.

Using the Poisson sum formula
equation (8) can also be expressed as the reciprocal space sum,
\begin{equation}
U=\frac{n^{2}\phi_{o}}{8\pi}\sum_{{\bf G}_{j}}\tilde{h}_{z}
\left({\bf G}_{j}\right),
\end{equation}
where $\tilde{h}_{z}({\bf G})$ is the Fourier transform of $h_{z}({\bf R})$ and
${\bf G}_{j}$ are the set of reciprocal lattice vectors.

The prescription for obtaining the equilibrium flux line lattice is
now as follows. First U is minimised with respect to the vortex
positions for a given density and orientation of the flux lattice with
respect to the $\hat{\bf c}$ axis. Finally, U is substituted into the
Gibbs free energy which is minimised as a function of the orientation
of the flux lattice.  Before U can be calculated, however, the
longitudinal component of the field distribution must be known.  This
is derived from the London theory in the next section.

\section{Anisotropic London Theory}
The high temperature superconductors may be viewed as a stack of
superconducting layers coupled via Josephson tunnelling. This is
conveniently described by the Lawrence-Doniach model\cite{ld}. The
Lawrence-Doniach model introduces four length scales: the in-plane
coherence length ($\xi_{ab}$) and penetration depth ($\lambda_{ab}$),
the distance between planes (s) and the Josephson length scale,
$\lambda_{J} =\xi_{ab}\sqrt{2/\rho}$, where $\rho$ is the
dimensionless Josephson coupling constant. $\lambda_{J}$ has the
physical significance that for distances from the vortex core of less
than $\lambda_{J}$ the phase differences between planes are large.
Conversely, for distances from the core of greater than $\lambda_{J}$
the phase differences between planes are small. In the latter case the
Josephson currents are small and in the extreme type II limit the
Lawrence-Doniach model becomes equivalent to the three dimensional
London theory with uniaxial anisotropy.

In YBa$_{2}$Cu$_{3}$O$_{7-\delta}$ $\lambda_{J}\approx$60{\AA}, which
for most typical field strengths is much smaller than the inter-vortex
spacing. Hence, the vortex-vortex interactions are accurately
described by London theory. Within the London approximation the basic
equation describing the magnetic field distribution for an isolated
vortex is
\begin{equation}
{\bf h}+\left(\nabla\times{\bf \Lambda}\cdot\nabla\times{\bf h}\right)
=\phi_{o}\delta\left({\bf r}\right){\bf \hat{z}},
\end{equation}
where ${\bf \Lambda}={\bf m}c^{2}/4\pi n_{s}e^{2}$, and {\bf m} is the
effective mass tensor. For uniaxial anisotropy {\bf $\Lambda$} has two
degenerate eigenvalues: $\Lambda_{a}$ associated with screening
currents flowing in the plane, and $\Lambda_{c}$ associated with
screening currents flowing along the $\hat{\bf c}$-axis. The
anisotropy ratio,
\begin{equation}
\gamma=\sqrt{\frac{\Lambda_{c}}{\Lambda_{a}}}=
\frac{\lambda_{c}}{\lambda_{a}}
\end{equation}
is defined by $\lambda_{J}/s$. Fourier transforming and inverting
equation (10) we obtain\cite{wbmg88}
\begin{equation}
{\bf \tilde{h}}({\bf k})=
\frac{1}{\left(1+\Lambda_{a}k^{2}\right)}
\left( {\bf \hat{z}}-
  \frac{\left(\Lambda_{c}-\Lambda_{a}\right)Q_{z}{\bf Q}}
{1+\Lambda_{a}k^{2}+\left(\Lambda_{c}-\Lambda_{a}\right)Q^{2}}\right)
\phi_{o},
\end{equation}
where ${\bf Q}={\bf k}\times \hat{\bf c}$.

The field component parallel to the vortex is
\begin{equation}
\tilde{h}_{z}({\bf k})=\frac{\phi_{o}(1+\lambda^{2}_{zz}k^{2})}{BC}
\end{equation}
where $B=1+\lambda^{2}_{zz}k^{2}_{x}+\lambda^{2}_{c}k^{2}_{y}$,
$C=1+\lambda^{2}_{a}k^{2}$, $k^{2}=k^{2}_{x}+k^{2}_{y}$ and
$\lambda^{2}_{zz}=\lambda^{2}_{a}\sin ^{2}\theta+\lambda^{2}_{c} \cos
^{2}
\theta$. Equation (13) is conveniently expressed as \cite{kog90}
\begin{equation}
\tilde{h}_{z}({\bf k})=\tilde{h}_{1}({\bf k})-\tilde{h}_{2}({\bf
k})=\frac{\phi_{o}}{\lambda^{2}_{a}}
\left(\frac{\lambda^{2}_{a}}{B}-\frac{\lambda^{2}_{zz}-
\lambda^{2}_{a}}{BC}
\right).
\end{equation}
The expression for $h_{z}({\bf r})$ is obtained from the inverse
Fourier transform of $\tilde{h}_{z}({\bf k})$ to produce
\begin{equation} h_{z}(r)=h_{1}(r)-h_{2}(r),
\end{equation}
where
\begin{eqnarray}
h_{1}(r) & = & \frac{\phi_{o}}{2\pi} \frac{\lambda_{zz}}
{\lambda^{2}_{a}\lambda_{c}}K_{o}(\rho_{o}),\nonumber\\
\rho_{o}^{2} & = & \frac{x^{2}}{\lambda^{2}_{zz}}+\frac{y^{2}}
{\lambda^{2}_{c}}
\end{eqnarray}
and
\begin{equation}
h_{2}(r) =  \frac{\phi_{o}}{2\pi} \left( \frac {\lambda^{2}_{zz}-
\lambda^{2}_{a}}{2\lambda^{2}_{a}} \right)
\int _{0}^{1} du \frac{\rho}{a(u)b(u)}K_{1}(\rho),
\end{equation}
with
\begin{equation}
\rho^{2}  =  \frac{x^{2}}{a^{2}(u)}+\frac{y^{2}}{b^{2}(u)},\nonumber
\end{equation}
\begin{equation}
a^{2}(u)  = \lambda^{2}_{zz}-(\lambda^{2}_{zz}-\lambda^{2}_{a})u
\end{equation}
and
\begin{equation}
b^{2}(u) = \lambda^{2}_{c}-(\lambda^{2}_{c}-\lambda^{2}_{a})u.
\end{equation}
$K_{0}$ and $K_{1}$ are the zeroth and first order Bessel functions,
respectively.

In isotropic superconductors the magnetic field due to a vortex is
 parallel to the vortex core and positive.  Hence, the Lorentz forces
 between a pair of vortices is centrally directed and repulsive.  In
 anisotropic superconductors, however, there are both longitudinal and
 transverse components to the vortex field. Furthermore, the
 longitudinal component becomes negative with minima situated at a
 distance $x_{min}$ either side of the core in the $\hat{\bf x}$
 direction ({\it i.e.} in the plane enclosing the $\hat{\bf c}$ axis
 and $\bf B$).  The Lorentz force between a pair of vortices therefore
 becomes attractive for certain relative orientations.  It is this
 attractive interaction which has led to the suggestion of the
 vortex-chain regime for flux densities where the average vortex
 spacing becomes comparable to $x_{min}$.  In figure 2 we plot
 $x_{min}$ as a function of the orientation of the vortex with respect
 to the $\hat{\bf c}$ axis for anisotropy ratios of 5.

\section{The Lattice Summation}

The interaction energy of a lattice of vortices is given in equations
(8) and (9). In this section we discuss the most efficient way of
performing these summations.

\subsection{Low Field Limit}

In the low field limit, defined by $L\gg\lambda$, where $L$ is the
average inter-vortex spacing, the real space sum converges rapidly.
The area of the unit cell, $s$, and hence $n$, is defined by,
\begin{equation}
s=\frac{1}{n}=\frac{\phi_{o}}{B}=\frac{L^{2}\sqrt{3}}{2}.
\end{equation}
The effect of the uniaxial anisotropy is to both lift the
orientational degeneracy of the hexagonal lattice and to cause a
uniaxial distortion of the equilateral triangles. A deformation of
$\beta_{x}$ in the $x$ direction and by $\beta_{y}$ in the $y$
direction will produce a family of isosceles triangles. For a fixed
flux density $s$ is constant, so that $\beta_{x} \beta_{y} =1$, and
hence we denote $\beta_{x}= 1/\beta_{y}=\beta$.  There are two
orientations of the lattice which are compatible with the uniaxial
symmetry. However, as shown in ref. [\onlinecite{kog88}] the
orientation of the lattice which minimises the energy for high flux
densities is that illustrated in figure 3.  This is also the
orientation which minimises the energy in the low density limit.

A general lattice vector corresponding to the unit cell shown in figure 3 is
\begin{eqnarray}
{\bf R}_{mn} & = & m{\bf a}_{1}+n{\bf a}_{2},\nonumber\\
\mbox{where}\
{\bf a}_{1} & = & L\beta \hat{\bf x},\nonumber\\
{\bf a}_{2} & = & L\left(\frac{\beta}{2} \hat{\bf x} + \frac{\sqrt{3}}{2\beta}
\hat{\bf y} \right),
\end{eqnarray}
and m and n are integers. The summations of $ h_{1}({\bf r})$ and $
h_{2}({\bf r})$ are  performed over ellipses
defined by the contours of $h_{1}
({\bf r})$ and $h_{2}({\bf r})$, respectively.

In the London approximation the vortex self-energy is divergent due to
the inability of the theory to describe the core adequately.  The
theory assumes a constant local magnetic field across the core, its
value being given at $r=\xi$. However, in anisotropic superconductors
the vortex core is not circular, but elliptical for fields oriented
away from $\hat{\bf c}$. Hence, an elliptical cut-off is required to
remove the divergence at small $r$, namely $\xi_{x}=\xi_{zz}$ and
$\xi_{y}=\xi_{a}$, where $\xi^{2}_{zz}=\xi^{2}
_{a}\cos^{2}\theta+\xi^{2}_{c}\sin^{2}\theta$ \cite{cutoff}.
($\xi_{a}$ and $\xi_{c}$ are the in-plane and out-off-plane coherence
lengths). The self-energy is therefore given by
\begin{equation}
U_{self-energy}=\frac{\phi_{o}}{2\pi} \frac{\lambda_{zz}}{\lambda^{2}_{a}
\lambda_{c}}K_{o}\left(\frac{1}{\kappa}\right)\nonumber\\
\end{equation}
where $\kappa=\lambda_{ab}/\xi_{c}=\lambda_{c}/\xi_{a}$.

\subsection{High Field Limit}

For high magnetic fields, $L \ll \lambda$, the real space sums converge slowly.
The summation of equation (8) can be more easily
evaluated in reciprocal space to obtain the lattice energy

\begin{equation}
U=\frac{Bn}{8\pi} \sum_{pq} \tilde{h}_{z}({\bf G}_{pq}),
\end{equation}
where $\tilde{h}_{z}({\bf G}_{pq})=\tilde{h}_{1}({\bf G}_{pq})-\tilde{h}_{2}
({\bf G}_{pq})$ and
\begin{equation}
\tilde{h}_{1}({\bf G}_{pq})=\frac{\phi_{o}\lambda^{2}_{zz}}{\lambda^{2}_{a}(
1+\lambda^{2}_{zz}G^{2}_{x}+\lambda^{2}_{c}G^{2}_{y})},
\end{equation}
\begin{equation}
\tilde{h}_{2}({\bf G}_{pq})=
\phi_{o}\left(\frac{\lambda^{2}_{zz}-\lambda^{2}_{a}}
{\lambda^{2}_{a}}\right)\int^{1}_{0}
\frac{du}{\left(1+a(u)^{2}G^{2}_{x}+b(u)^{2}G^{2}_{y}
\right)^{2}}.
\end{equation}
The coefficients $a(u)$ and $b(u)$ are given in equations (19) and (20).
{\bf G}$_{pq}$ describes the set of reciprocal lattice vectors,
\begin{eqnarray}
{\bf G}_{pq} & = & p{\bf b}_{1}+q{\bf b}_{2},\nonumber\\
\mbox{where}\
{\bf b}_{1} & = & G_{o}\left( \frac{\sqrt{3}}{2\beta}\hat{\bf x}-\frac{\beta}
{2}
\hat{\bf y}\right),\nonumber\\
\mbox{and}\
{\bf b}_{2} & = & G_{o}\beta \hat{\bf y},
\end{eqnarray}
with $G_{o}=2\pi L/s$, and p and q are integers.

The sum of $\tilde{h}_{2}({\bf G}_{pq})$ converges rapidly. However,
the sum of $\tilde{h}_{1}({\bf G}_{pq})$ is logrithmically diverging
because of the $1/G^{2}$ behaviour. Physically this corresponds to the
self-energy of the vortex core. There are various ways to deal with
this divergence. One method \cite{cutoff} is to use a Gaussian cut-off
in the reciprocal space sum. Another is to use the Ewald summation
method, originally used by Fetter \cite{fetter} for type II
superconductors. The advantage of the latter method is that it
explicitly removes the formally diverging self energy term, so we
adopt it in this paper.

The details of the extension of Fetter's method to anisotropic
superconductors are shown in the appendix. Here we simply quote the
result for the total energy per unit volume, excluding the
self-energy, as
\begin{eqnarray}
U
& = &
\frac{Bn\phi_{o}}{8\pi}\frac{\lambda^{2}_{zz}}{\lambda^{2}_{a}}\left(1-
\alpha^{2} +
\sum_{\tilde{\bf G}\neq0}\left[\frac{\exp\left(-\alpha^{2}\tilde{\bf G}^{2}
\right)}
{\tilde{\bf G}^{2}}-\frac{1}{\tilde{\bf G}^{2}\left(1+\tilde{\bf G}^{2}\right)}
\right]+
\frac{1}{4\pi n\lambda_{zz}\lambda_{c}}\sum_{\tilde{\bf R}_{i}\neq 0}
E_{1}\left(\frac{\tilde{\bf R}_{i}^{2}}{4\alpha^{2}}\right)\right)\nonumber\\
& - &
\frac{Bn\phi_{o}}{8\pi}\left(\frac{\lambda^{2}_{zz}-\lambda^{2}_{a}}
{\lambda^{2}_{a}}\right)\sum_{\tilde{\bf G}}\int^{1}_{0}
\frac{du}{\left(1+a^{2}G^{2}_{x}+b^{2}G^{2}_{y}\right)^{2}}
\end{eqnarray}
where $\alpha^{2}=1/4\pi n$, $\tilde{\bf G}^{2}=\lambda^{2}_{zz}G^{2}_{x}
+\lambda^{2}_{c}G^{2}_{y}$, $\tilde{\bf R}^{2}=({\bf R}_{x}/\lambda_{zz})^{2}
+({\bf R}_{y}/\lambda_{c})^{2}$ and $E_{1}$ is the exponential integral.

The vortex self-energy (obtained in the limit $\tilde{\bf R}_{i}\rightarrow0$)
 is not included in this
summation, but can be evaluated as before by employing an elliptical cut-off
for the vortex core, namely $x\rightarrow\xi_{x}=\xi_{zz}$ and $y\rightarrow
\xi_{y}=\xi_{c}$. Hence, with the substitution $\tilde{\bf R}_{i}=1/\kappa$,
the self-energy term becomes
\begin{equation}
U_{self-energy}=\frac{B\phi_{o}\lambda_{zz}}{32\pi^{2}\lambda^{2}_{a}\lambda_
{c}}
E_{1}\left(\frac{\pi n}{\kappa^{2}} \right)
\end{equation}

\section{Results and Discussion}

We now turn to a discussion of our predictions for the lattice
 parameters, and their comparsion with experiment. Figure 4a shows our
 predictions and the experimental values from [\onlinecite{gammel92}]
 of the intra-chain (D) and inter-chain (C) distances {\it when
 projected onto the a-b plane}, versus the normal component of the
 applied field.  The applied field is at an angle of $40^o$ to the
 $\hat{\bf c}$ axis. We have taken the anisotropy parameter $\gamma$
 as 5, the in-plane penetration depth $\lambda _{a}$ as 1 413{\AA},
 and the in-plane coherence length $\xi_{a}$ as 16{\AA}. For this
 choice of parameters there is very good agreement with experiment.
 The results for the same parameters are shown in figure 4b for an
 applied field at $70^o$ to the $\hat{\bf c}$ axis, where the fit is
 less good. A possible explanation for this might be due to a
 misalignment of the crystal.  The inset of figure 4b shows the flux
 line lattice orientation with respect to the $\hat{\bf c}$-axis as a
 function of the applied magnetic field. For low magnetic fields the
 flux lattice is oriented nearly parallel to the $\hat{\bf c}$ axis,
 slowly aligning with the applied field direction with increasing
 magnetic intensity.

To show that the vortex-chain structure has indeed been observed in
reference [\onlinecite{gammel92}] we must take into account the
effects of the rotation of the flux lattice as a function of applied
field. This has two consequences: first, the position of the minimum
in the vortex-vortex interaction, $x_{min}$, will change, as it is
angular dependent. This angular dependence is shown in figure
2. Second, the magnetic flux density in the plane normal to the flux
lattice is not linearly related to the applied field, but given by
equation (3).  We therefore scale the results in the following
way. The intra-chain distance {\it in the plane normal to the flux
lattice}, $d=D\cos\theta$, is divided by $x_{min}$ and plotted against
the inverse flux density, $1/B$.  In the vortex-chain regime this
quantity should be a constant. To preserve areas the inter-chain
distance is multiplied by $x_{min}$, and should be inversely
proportional to the flux density. Figures 5a and 5b shows the log-log
plot of $d/x_{min}$ and $c\times x_{min}$ against $1/B$ for the
applied field orientations of 40$^{o}$ and 70$^{o}$, respectively.  We
have also plotted the experimental points scaled in the same way.
Evidently, for low fields, both the theoretical and experimental
results are in close agreement with the vortex-chain predictions, with
$d/x_{min}$ roughly unity and almost independent of B.

At higher fields, scaling of the unit cell parameters vary with the
magnetic flux density as $\sim 1/B^{1/2}$, as expected. The
anisotropic distortion of the hexagonal lattice agrees with the
perturbation expansion of [\onlinecite{kog88}] with,
\begin{equation}
\beta=\left( {\frac{ \sin^2\theta + \gamma^2\cos^2\theta}{\gamma^2}}\right)^
{1/4}.
\end{equation}
At high fields the flux lattice is parallel to the applied field, so
$\theta =\phi$.

As a final comparison to experiment we calculate the value of the
intra-chain distance on the surface of the sample as a function of the
orientation of the applied field for a {\it fixed normal component} of
12 Oe.  This is shown in figure 6 with the experimental values of
[\onlinecite{gammel92}] for comparison.  Again, there is reasonable
agreement for this choice of parameters.

\section{Conclusions}
\noindent

In summary, we have calculated the equilibrium flux line lattice in
planar crystals of YBa$_{2}$Cu$_{3}$O$_{7-\delta}$ using three
dimensional anisotropic London theory.  By taking into account
demagnetisation effects, which cause the flux line lattice to orient
away from the applied field towards the crystalline axis, we are able
to show that the low field Bitter pattern experiments of Gammel {\it
et al.}\cite{gammel92} demonstrate the existence of the
``vortex-chain'' state.  In this state the intra-chain distance is
independent of flux density, whereas the inter-chain distance scales
as the inverse flux density, 1/B.  We also demonstrate that as the
field strength is increased there is a smooth cross-over to the
distorted hexagonal lattice in which the inter-vortex spacings scale
uniformly as $1/B^{1/2}$.

The agreements between theory and experiment are found for an
anisotropy ratio $\gamma$ of 5 and the in-plane penetration depth
$\lambda_{a}$ of 1 413{\AA}.  This value of the penetration depth is
the zero temperature penetration depth, and not the value of the
penetration depth at the irreversibility line quoted for twinned
samples.  Since the samples used in ref [\onlinecite{gammel92}] are
relatively twin-free, consisting of untwinned regions of at least
100$\mu$m square, we deduce that locally the lattice is not frozen but
has assumed its zero temperature equilibrium configurations.
\vspace {0.5 truein}

\begin{center}

Acknowledgements

\end{center}

We thank E. M. Forgan, P. de Groot and N. Wilkin for helpful
discussions. M.H. is supported by a University of Sheffield
Scholarship.  W.B. acknowledges a grant from the University of
Sheffield Research Fund. Part of this work was funded by the Ministry
of Defence, agreement no. 2031/153/csm.

\pagebreak

\appendix
\section{}

In this appendix we use the Ewald summation method to derive the
interaction energy of a lattice of vortices in the high density limit.

The reciprocal space sum

\begin{equation}
S=\sum_{\mbox{all} {\bf \tilde{G}}}\frac{1}{1+\tilde{\bf G}^{2}}
\end{equation}
diverges logarithmically for large $\tilde G$, where $\tilde{\bf G}^{2}=
\lambda^{2}_{zz}G^{2}_{x}+\lambda^{2}_{c}G^{2}_{y}$.

(A1) may be written as
\begin{equation}
S=1+\sum_{\tilde{\bf G}\neq0}\frac{1}{\tilde{\bf G}^{2}}-
\sum_{\tilde{\bf G}\neq 0} \frac{1}{\tilde{\bf G}^{2}\left(
1+\tilde{\bf G}^{2}\right)},
\end{equation}
where the third term converges, leaving the second
term to consider.  Using the identity
\begin{equation}
\frac{1}{\tilde{\bf G}^{2}} =
2\int^{\infty}_{0} \xi \exp\left(-\xi^{2}\tilde{\bf G}^{2}\right)d\xi,
\end{equation}
and splitting the integral into two parts, we obtain
\begin{eqnarray}
\sum_{\tilde{\bf G}\neq0}\frac{1}{\tilde{\bf G}^{2}}
& = &
\sum_{\tilde{\bf G}\neq0}\left[ 2\int^{\alpha}_{0} \xi \exp\left(
-\xi^{2}\tilde{\bf G}^{2}\right)d\xi
+ \int_{\alpha}^{\infty} \xi \exp\left(
-\xi^{2}\tilde{\bf G}^{2}\right)d\xi \right] \nonumber\\
& = &
\sum_{\tilde{\bf G}\neq0}\left[ 2\int^{\alpha}_{0} \xi \exp\left(
-\xi^{2}\tilde{\bf G}^{2}\right)d\xi
+ \frac{\exp{\left(-\alpha^{2}\tilde{\bf G}^{2}\right)}}
{\tilde{\bf G}^{2}} \right]
\end{eqnarray}
where the choice of $\alpha=(4\pi n)^{-1/2}$ maximises the rate of convergence
and $n$ is the real space density of lattice points. The second
term of (A4) now has a Gaussian cut-off and converges easily. However, the
logarithmic divergence has to be extracted from the first term.
Writing the first term of (A4) as a sum over all $\tilde{\bf G}$,
\begin{displaymath}
T= \sum_{\mbox{all}\tilde{\bf G}}
2\int^{\alpha}_{0} \xi \exp\left(-\xi^{2}\tilde{\bf G}^{2}\right)d\xi
-2\int^{\alpha}_{0}\xi\ d\xi
\end{displaymath}
and using the Poisson sum formula, $n\sum_{\bf G} \tilde F(\bf G) = \sum_{\bf
R} F({\bf R})$, we obtain
\begin{equation}
T=\frac{1}{4\pi n\lambda_{zz}\lambda_{c}}\sum_{\mbox{all} \tilde{\bf R}_{i}}
2 \int^{\alpha}_{0}\exp\left(-\frac{\tilde{\bf R}^{2}}{4\xi^{2}}\right)
\frac{\mbox{d}\xi}{\xi}-\alpha^{2},
\end{equation}
where $\tilde{\bf R}^{2}=({\bf R}_{x}/\lambda_{zz})^{2}+({\bf R}_{y}/
\lambda_{c})^{2}$.

Finally, collecting terms,
\begin{eqnarray}
S=1
& + &
\sum_{\tilde{\bf G}\neq0}\left[\frac{\exp\left(-\alpha^{2}\tilde{\bf G}^{2}
\right)}
{\tilde{\bf G}^{2}}-\frac{1}{\tilde{\bf G}^{2}\left(1+\tilde{\bf G}^{2}\right)}
\right]\nonumber \\
& + &
\frac{1}{4\pi n\lambda_{zz}\lambda_{c}}\sum_{\tilde{\bf R}_{i}}E_{1}
\left(\frac{\tilde{\bf R}^{2}_{i}}{4\alpha^{2}}\right)-\alpha^{2},
\end{eqnarray}
where the exponential integral is
\begin{displaymath}
E_{1}\left(x\right)=\int^{\infty}_{x}\frac{\exp(-t)}{t}dt.
\end{displaymath}
The logarithmically diverging contribution comes from the term $\tilde{\bf R}
_{i}\rightarrow0$ in (A6).

\pagebreak

\begin{center}{\bf  Figure Captions}\end{center}

Figure 1. The slab geometry used in the decoration experiments, with
the $\hat{\bf c}$-axis perpendicular to the surface of the slab.  The
($\hat{\bf X},\hat{\bf Y},\hat{\bf Z}$) axes define the
crystallographic axes with $\hat{\bf Z}$ being parallel to $\hat{\bf
c}$ and ($\hat{\bf X}-\hat{\bf Y}$) lying in the ($\hat{\bf
a}-\hat{\bf b}$) plane.  The external field, ${\bf H}_{a}$, is applied
at an angle $\phi$ to $\hat{\bf c}$, producing a vortex lattice of
magnetic flux density, {\bf B}, inclined at an angle $\theta$ to
$\hat{\bf c}$. This vortex lattice is described by a general
coordinate axes ($\hat{\bf x},\hat{\bf y},\hat{\bf z}$) by a rotation
of $\theta$ about the $\hat{\bf Y}$-axis such that $\hat{\bf c}$,
$\hat{\bf x}$ and $\hat{\bf z}$ are coplanar and {\bf B} is parallel
to $\hat{\bf z}$.

Figure 2.  The position of minimum of the inter-vortex potential,
$x_{min}$, in units of $\lambda = \sqrt{\lambda_a \lambda_c}$ versus
the orientation of the vortex for $\gamma =5$.

Figure 3. (a) The isotropic flux line lattice is described by
equilateral triangles of side L. (b) A deformation of $\beta$ in the
$x$ direction and by $1/\beta$ in the $y$ direction will produce a
family of isosceles triangles that describe the anisotropic flux line
lattice.

Figure 4(a). Intra-chain distance (D): theory, solid line; experiment,
open diamonds, and inter-chain distance (C): theory, dashed line;
experiment, filled diamonds, projected onto the a-b plane. Applied
field orientation of 40$^{o}$, in-plane penetration depth
$\lambda_{a}$ of 1 413{\AA}, in-plane coherence length $\xi_{a}$ of
16{\AA} and anisotropy ratio $\gamma$ of 5. The inset shows the vortex
chain.

Figure 4(b). The same as fig 4(a) with an applied field orientation of
70$^{o}$.  The inset shows the orientation of the flux line lattice as
a function of the applied field: solid line - $\phi$=40$^{o}$ and
dotted line - $\phi$=70$^{o}$.

Figure 5(a). Log-log plot of intra-chain distance divided by
$x_{min}$: theory, solid line; experiment, open diamonds and
inter-chain distance multiplied by $x_{min}$: theory, dashed line;
experiment, filled diamonds, in the plane normal to the flux
lattice. Applied field at 40$^{o}$.

Figure 5(b).  The same as fig 5(a) with the applied field at 70$^{o}$.

Figure 6. The intra-chain distance on the surface of the sample as a
function of on the applied field orientation for a fixed normal
component of 12 Oe, open squares - experiment.

\end {document}